\documentclass[11pt,twoside]{article}

\usepackage{asp2004}
\usepackage{epsf}
\usepackage{psfig}
\usepackage{lscape}
\usepackage{comment}

\begin{document}
\title{Phenomenological Modeling of the FIR--Radio Correlation within
  Nearby Star-Forming Galaxies}   
\author{E.J. Murphy$^{1}$, G. Helou$^{2}$, L. Armus$^{2}$,
  R. Braun$^{3}$, J.D.P. Kenney$^{1}$, and the SINGS team} 
\affil{$^{1}$Yale University; $^{2}${\it Spitzer} Science
  Center/Caltech; $^{3}$ASTRON} 

\begin{abstract} 
We present an analysis of the far-infrared (FIR)-radio correlation
within a group of nearby star-forming galaxy disks observed as part of
the {\it Spitzer} Infrared Nearby Galaxies Survey (SINGS).
In our study we critically test a phenomenological model for the
FIR-radio correlation which describes the radio image as a smeared
version of the infrared image.
The physical basis for this model is that cosmic-ray electrons (CR
electrons) will diffuse significant distances from their originating
sources before decaying by synchrotron emission.
We find that this description generally works well, improving the
correlation between the radio and infrared images of our four sample
galaxies by an average factor of $\sim$1.6.
We also find that the best-fit smearing kernels seem to show a
dependence on the ongoing star formation activity within each disk.
Galaxies having lower star formation activity (NGC~2403 and NGC~3031)
are best-fit using larger smearing kernels than galaxies with more
active star-forming disks (NGC~5194 and NGC~6946). 
We attribute this trend to be due to a recent deficit of CR electron
injection into the interstellar medium of galaxies with lower star
formation activity throughout their disks.  

\end{abstract}


\section{Introduction}
The near-universality of the far-infrared (FIR)--radio correlation
lends it to be a critical tool for constraining the physical processes
that shape galaxies.
While the FIR-radio correlation can teach us how the global parameters
of galaxies work with one another to keep the integrated thermal dust
emission in constant proportion with the non-thermal radio synchrotron
emission, 
resolved studies can help us understand the link between relativistic
and non-relativistic phases of the interstellar medium (ISM) on a much
more intimate level. 

To date, nearly all of our observational knowledge about the
relativistic phase (cosmic-ray component) of the ISM comes from direct
measurements of cosmic-ray (CR) nuclei in the solar neighborhood.
These observational characteristics of the CR component of the Galaxy
are normally considered to be applicable for all galaxies.
The only observational constraints on CR physics in other galaxies
comes from multi-wavelength radio observations which are able to
probe the CR electron component of galaxies (eg. \citet{ul96}).
However, these data alone only describe the present distribution
of CR electrons without providing insight on the CR electron source
distribution and, consequently, the diffusion and propagation history
of the CR electrons. 

Previous studies of the FIR-radio correlation using {\it IRAS} data
have tried to explain it using a phenomenological model in which the
radio image appears as a smeared version of the infrared image
\citep{bh90, mh98}.
The physical backing of this picture arises from the fact that CR
electrons diffuse much larger distances ($\sim$1-2~kpc) compared to
the mean free path of dust heating UV photons ($\sim$100~pc). 
\citet{ejm06} have recently shown that the correlation between
the FIR and radio emission within galaxies on sub-kpc scales displays
a non-linearity with FIR/radio ratio increasing with FIR surface
brightness.  
This non-linearity can be attributed to the diffusion of CR electrons
away from star-formation sites in agreement with the 'image-smearing'
picture.
In this conference proceedings, we highlight the results of the
phenomenological 'image-smearing' model when applied to a sample of
four nearby star-forming galaxies (NGC~2403, NGC~3031, NGC~5194, and
NGC~6946) using {\it Spitzer} MIPS and WSRT radio continuum data,
observed as part of the {\it Spitzer} Nearby Galaxies Survey (SINGS)
\citep{rk03}. 
These galaxies display a range in their star formation activity and
Hubble types, but are at distances that allow us to study the
correlation on sub-kpc scale.

\begin{figure}[!ht]
\plotone{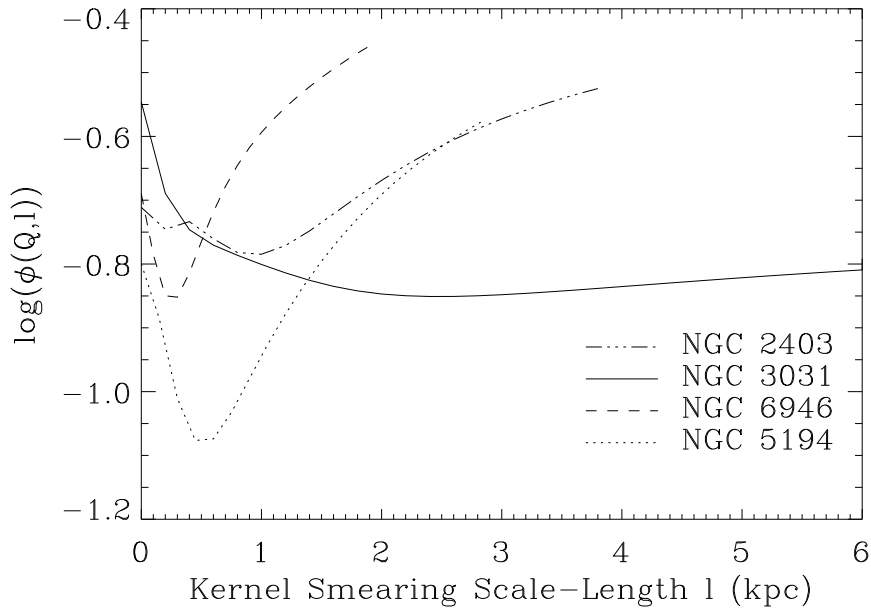}
\caption{Residuals between the smeared 70~$\micron$ and radio maps are
  plotted as a function of the smearing kernel scale-length for each
  of the four sample galaxies.  Notice the different behavior among the
  residual curves.  
  Residual curves for galaxies with higher star formation activity
  (NGC~5194 and NGC~6946) display narrow troughs with well defined
  minima.  
  The less active star-forming galaxies (NGC~2403 and NGC~3031) have
  much broader troughs with less well defined minima. 
  NGC~2403 even shows evidence for two minima at $\sim$0.2 and
  $\sim$1.0~kpc which may be the result of a superposition of two
  distinct CR electron populations having significantly different
  ages.  
  \label{rescurves}} 
\end{figure}

\section{Results}
In Figure \ref{rescurves} we plot the residuals between the smoothed
70~$\micron$ and radio maps against the scale-length ($l$) of the
smearing kernel.
The 70~$\micron$ images are smoothed by exponential kernels projected
in the plane of the sky which take the form \(\kappa({\bf r}) =
e^{-r/l}\).
A comparison of the residual curves among the four sample galaxies
shows a range in behavior and a factor of ~$\sim$1.6 mean improvement
in the correlation.
NGC~5194 and NGC~6946 both have well behaved residual curves,
displaying rather narrow troughs with well defined minima
corresponding to smearing scale-lengths less than 1~kpc.
NGC~2403 and NGC~3031, on the other hand, exhibit different behavior. 
The residual curve for NGC~3031 is very shallow and has a minimum
centered around $\sim$2.5~kpc that is not well constrained.
In the case of NGC~2403, we see evidence for two minima in its
residual curve.
The first minimum corresponds to a smearing scale-length of
$\sim$0.2~kpc, while the second, and deeper of the two, occurs for a
scale-length around $\sim$1.0~kpc.

\begin{figure}[!ht]
  \resizebox{12.5cm}{!}{
    {\plotone{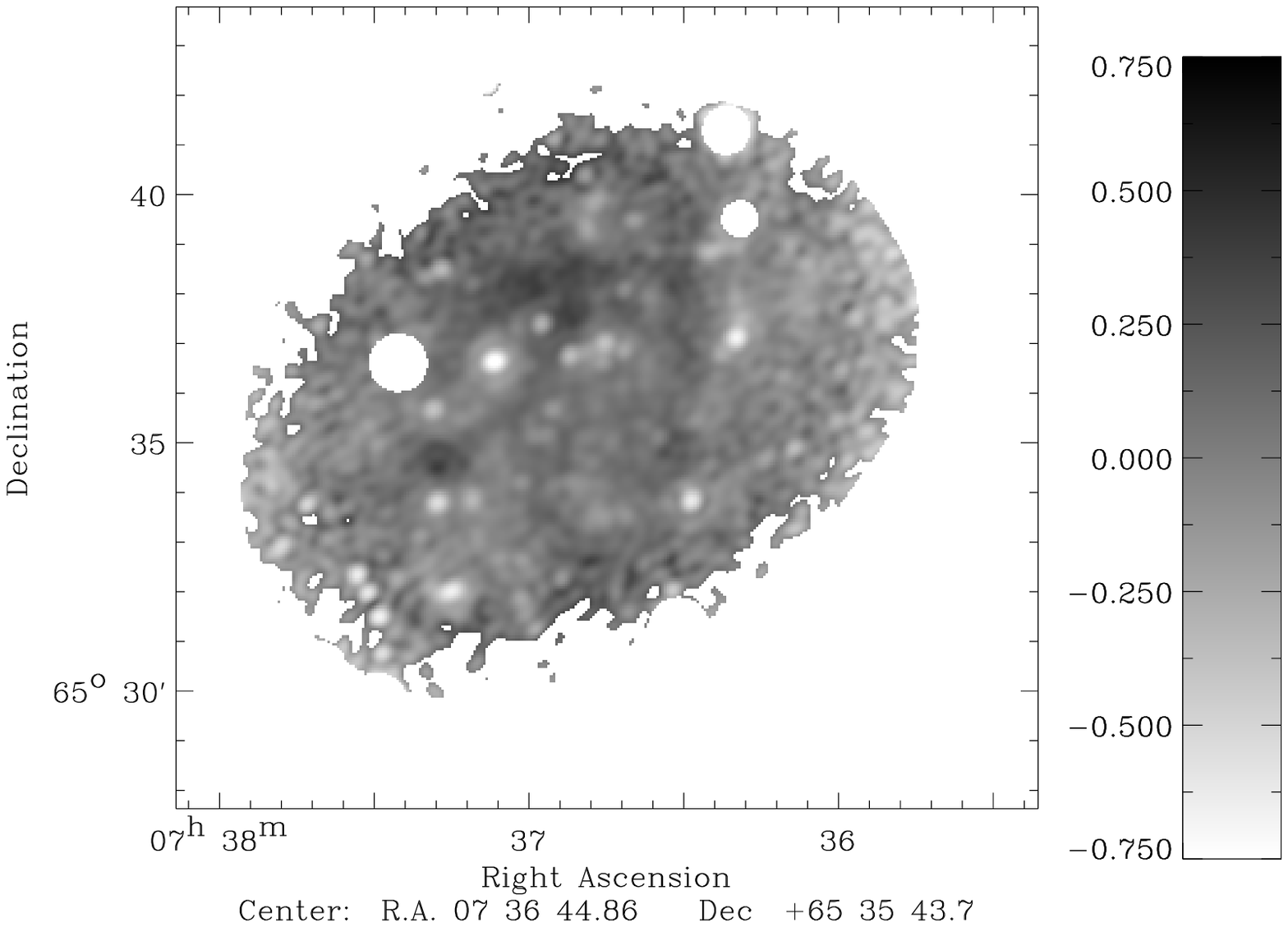}
      \plotone{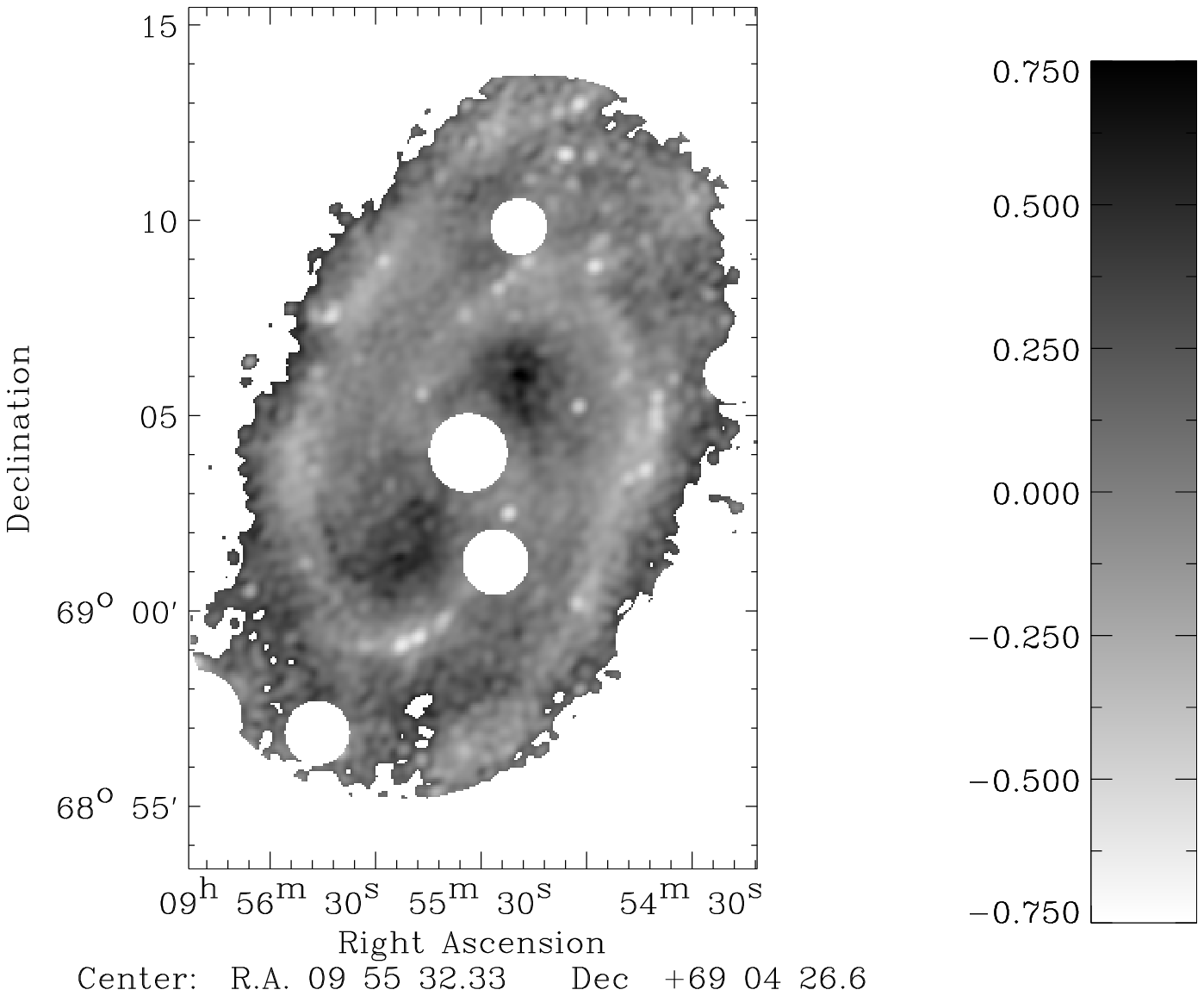}}}
  \vskip 1cm
  \resizebox{12.5cm}{!}{
    {\plotone{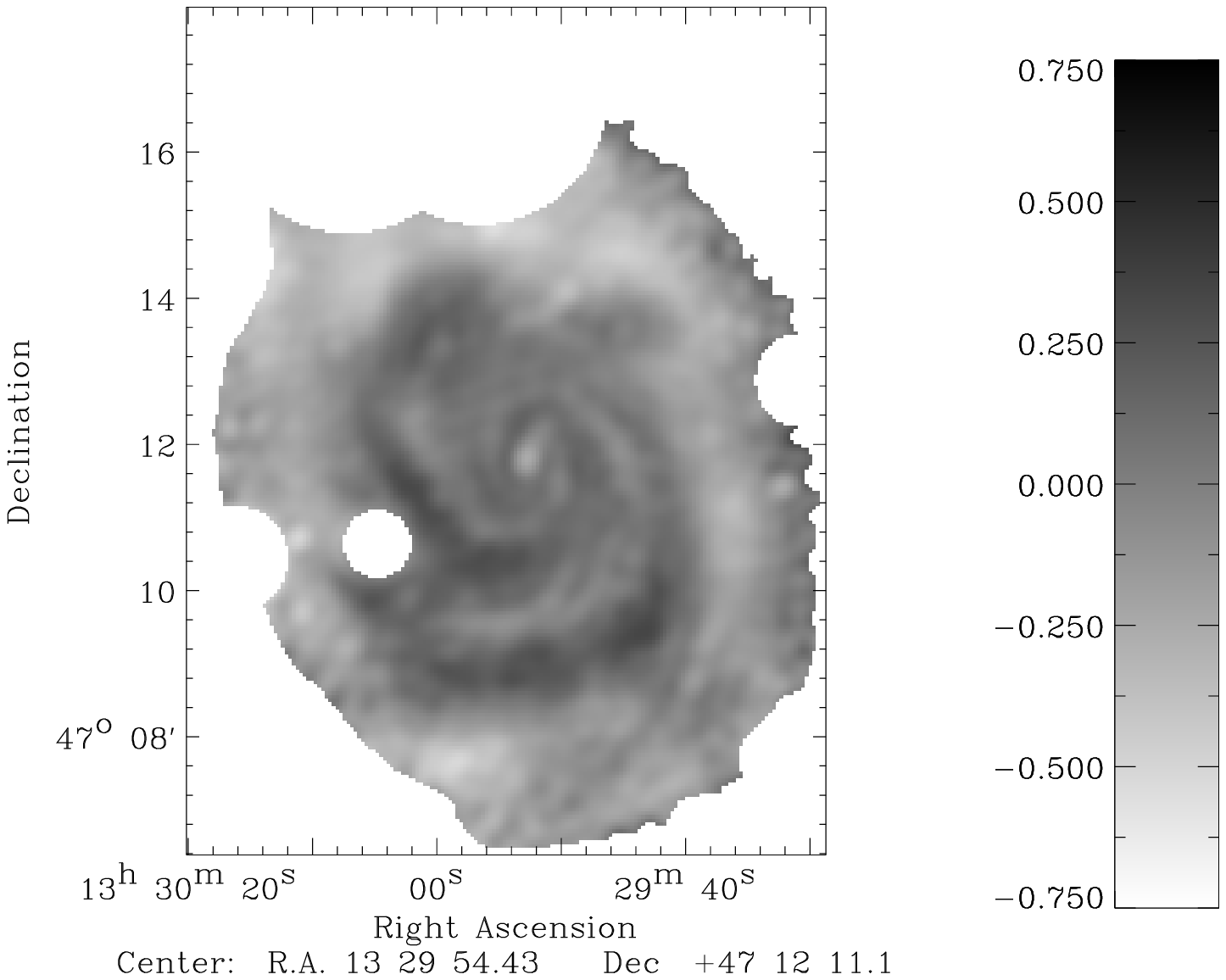}
      \plotone{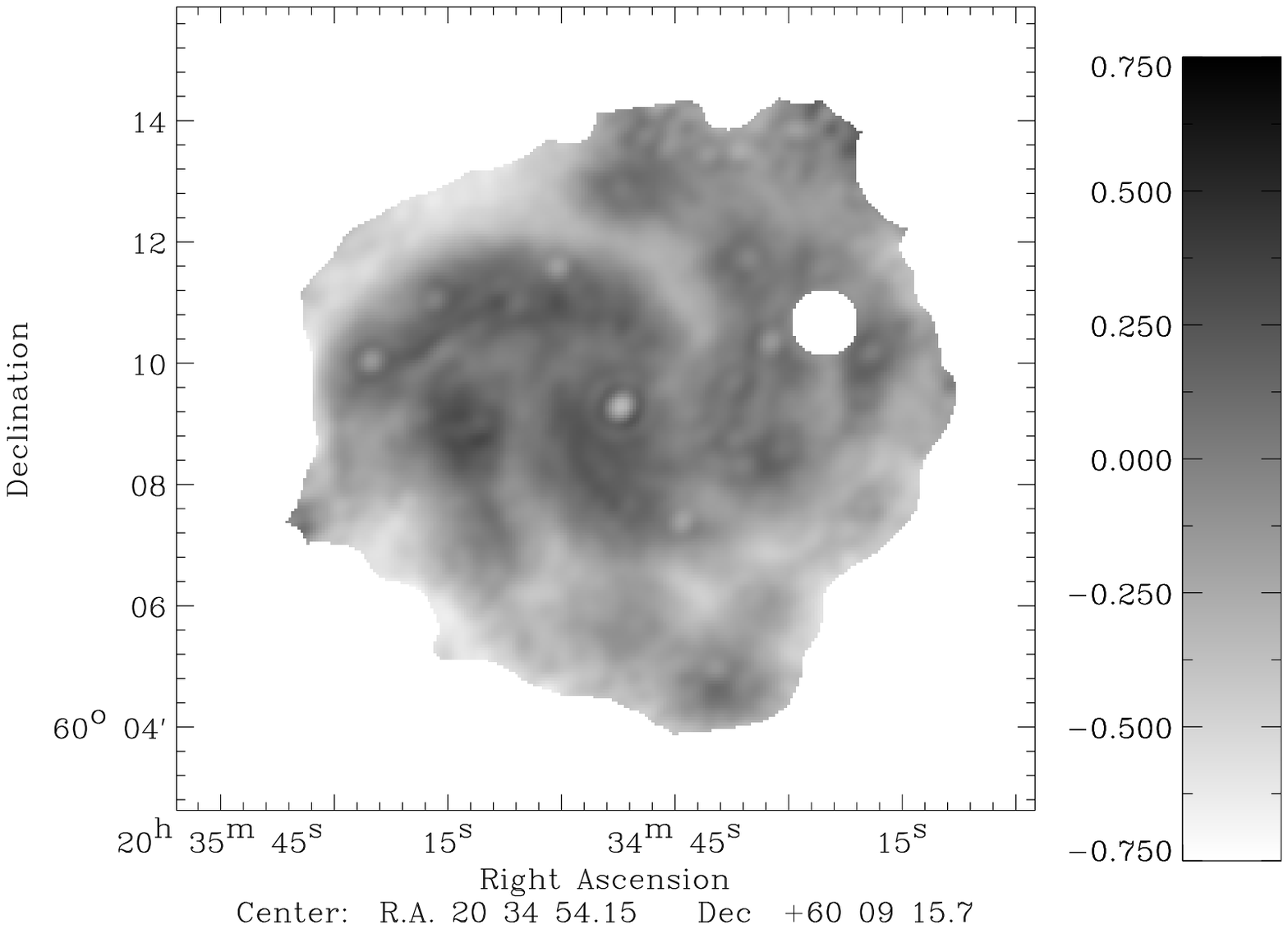}}}
  \vskip .5cm
  \caption{Residual images after subtracting the observed radio maps
    from the smeared 70~$\micron$ images for each galaxy using the
    best-fit exponential kernel oriented in the plane of the sky.
    Notice the non-noiselike residuals left behind which seem to vary
    from one galaxy to the next.
    The residual maps for galaxies with higher star formation activity
    (NGC~5194 and NGC~6946) display infrared excess associated with
    star formation sites.
    Galaxies with lower star formation activity (NGC~2403 and
    NGC~3031) display an opposite trend in which star formation sites
    exhibit radio excesses due to a large smearing kernel.
    \label{resmaps}}
\end{figure}

In Figure \ref{resmaps} we plot the residual maps associated with the
best-fit smearing kernel. 
For each galaxy we find non-noiselike residuals displaying
organized structure co-spatial with areas of star formation. 
We also find distinct and opposite behavior in the residual maps of
NGC~2403 and NGC~3031 when compared with those for NGC~5194 and
NGC~6946.    
While NGC~2403 and NGC~3031 display radio excesses in their residual
maps associated with sites of star formation, NGC~5194 and NGC~6946
exhibit infrared excesses around star-forming regions.
This is not surprising since galaxies best fit with large smearing
kernels have increased amounts of flux redistributed away from bright
star-forming regions which will result in radio excesses for these
locations in the residual maps. 

\section{Discussion}
Using new high resolution {\it Spitzer} imaging at 70~$\micron$ for a
sample of 4 galaxies, 
we have applied a phenomenological model for the FIR-radio
correlation, which describes the distribution of radio emission as a
smoother version of the infrared flux distribution.
We find that the model fitting parameters seem to differ depending on
the amount of ongoing star formation activity within each galaxy.

While we see dramatically different residual behavior for NGC~5194 and
NGC~6946 compared to NGC~2403 and NGC~3031, we also find that these
pairs of galaxies have quite different ongoing star formation within
their disks.  
Both NGC~5194 and NGC~6946 have star-formation rates (SFRs) which are
roughly a factor of $\sim$6 higher than NGC~2403 and NGC~3031
\citep{ejm06}.
We attribute the differences in the residuals among the four galaxies
to be due to timescale effects.  
In the less active star-forming galaxies there has likely been a
deficit of recent CR electron injection into their ISM leaving the
diffuse disk as the dominant morphological structure.
Accordingly, in the more active star-forming galaxies, CR electrons
are likely being injected into the ISM at a much higher rate, thereby
keeping both the mean age of the CR electron distribution young and
their mean diffusion lengths short.  
We speculate the double minima behavior for the residual curve of
NGC~2403 to be due to a superposition of two distinct CR electron
populations having significantly different ages.  


\acknowledgements 
E.J.M. would like to acknowledge support provided by the {\it Spitzer}
Science Center Visiting Graduate Student program. 
As part of the {\it Spitzer} Space Telescope Legacy Science Program,
support was provided by NASA through Contract Number 1224769 issued
by the Jet Propulsion Laboratory, California Institute of Technology
under NASA contract 1407.


\end{document}